\documentclass{article}
\usepackage{geometry}  
\usepackage{graphicx}
\usepackage[round]{natbib}
\usepackage{graphicx}
\usepackage{amssymb}
\usepackage{epstopdf}
 \usepackage{siunitx} 
\title{Are renewable energies on a sustained path? \\
 Analysis of selected case-studies from the pre-pandemic-era}
\author{Alessandro Bessi  \\
	Department of Economics and Management, University of Pisa \\
	\and 
	Mariangela Guidolin \\
	Department of Statistical Sciences, University of Padua\\
	\and Piero Manfredi \\
	Department of Economics and Management, University of Pisa}

\begin{document}
\maketitle


\begin{abstract} 
Provided widespread vaccination will bring the COVID-19 pandemic under full control worldwide, the contrast to climate change and the energy transition as
one of its main actions will return at the top of national and international policy agendas.\\
This paper employs multivariate diffusion models to investigate and quantitatively assess the competitive power of renewable energy technologies and their perspectives along the invoked energy transition. The study was conducted for the period 1965-2019 on a number of selected case studies, that were considered critically representative of the current transition process in view of their energy and political context. 
The dynamic relationship between renewable technologies and natural gas has been at the core of the analysis, trying to establish whether gas could be considered as a \textit{bridging technology} or a \textit{lock-in}. 
The main findings show that in all the analyzed countries RETs have exerted a strongly competitive effect towards gas. In most cases, gas is found to have a bridging role, aiding the uptake of renewables. 
\end{abstract}

\section{Introduction}
\label{sect:1}

Provided widespread vaccination will bring the COVID-19 pandemic under full control worldwide, the contrast to climate change and the energy transition as one of its main actions will return at the top of national and international policy agendas. Despite most of the green actions planned in the pre-pandemic epoch, such as the ``Green New Deal'' (\cite{mastini:21}), envisaged substantial resources allocation for green investments, and  all ``Recovery plans'' foresee a dramatic strengthening of such actions, the future perspectives of the renewable energy technologies, RETs, need to be investigated in light of the complex dynamics characterizing energy markets. \\
Several recent research contributions are trying to identify useful lessons from the COVID-19 experience for the future planning and management of energy systems. \cite{hoang:21} and \cite{jiang:21} identified emerging opportunities, challenges and policy indications learned from the pandemic, while \cite{heffron:21} stressed the importance of energy resilience as countries face the
triple challenge of the COVID-19 health crisis, the consequent economic crisis, and the climate crisis. \\
As an interesting insight in this debate, \cite{IEA:21} defined renewables as the `success story' of the COVID-19 era, with solar PV and wind expected to contribute two-thirds of renewables’ growth. Indeed, the share of renewables in electricity generation is projected to increase to almost 30\% in 2021, the highest share since the beginning of the Industrial Revolution. This study also reported that global natural gas consumption declined in 2020, although with a minor decrease with respect to coal and oil, therefore proving a stronger relative resilience. According to IEA, this resilience can be partly explained by fuel switching in electricity generation, a particularly remarkable phenomenon in the United States.\\
In order to quantitatively assess the competitive power of RETs and their perspectives during the incoming transition,
this paper employs multivariate diffusion models to describe the past and current trend of energy markets on a few selected case studies, considered critically representative of the current transition process in view of their energy and political context. 
Specifically, the top leading countries for renewable energy consumption in 2019, according to the \cite{BP:20}, were considered, namely China, US, Germany, Brazil, India, Japan, UK, Spain, Italy and France. 
Additionally, we included in the analysis Denmark and Turkey. The former was added as a paradigmatic example of an industrialized country that heavily invested on RETs, while the latter is representative of an emerging economy with a robust positive trend in the sector. \\
The dynamic relationship between renewable technologies and natural gas was at the core of the analysis, since gas has been typically defined as a `transition mean', given its reduced emissions compared to other fossil fuels used for electricity provision, such as coal. As suggested by \cite{guidolin:19} gas can, depending on the energy setting and its determinants, play an ambivalent role, either stimulating or hindering the adoption of RETs. \cite{gursan:20} proposed a thorough discussion of this aspect, by reviewing the most recent literature on energy transition and the role of gas, and highlighted that gas may be considered either a \textit{bridging technology} or a \textit{lock-in}, depending on the dynamic relationships considered. Indeed, gas may have a direct positive effect on the growth of renewables, but also an indirect negative effect, a \textit{crowd-out effect}, for which a bridging technology subtracts investments to the emerging technology. 
In this perspective, the authors suggested that to avoid lock-ins, it is crucial to understand the trajectories of each technology and the interdependencies existing between them. 

Consistently with our main goals, the paper uses an established bivariate diffusion model namely the UCRCD model proposed in \cite{guseo:14}, to understand whether and to which extent renewable energy had a competitive power over traditional energy sources for electricity provision. \\
The rest of the paper is organized as follows: Section \ref{sect:2} describes some background literature in innovation diffusion modeling applied to energy markets, Section \ref{sect:3} illustrates the methodological approach for the analysis, Section \ref{sect:4} presents the results of modeling and discusses the main implications of them in the energy context. Section \ref{sect:5} is left for some concluding remarks and final discussion, with some proposals for future research.

\section{Background literature}
\label{sect:2}
Since widespread adoption of RETs plays a crucial role in the energy transition, the ability to understand and effectively model their diffusion is essential to orient policy decisions, set achievable targets and stimulate virtuous market mechanisms. 
In recent years, the literature concerning RETs has grown considerably and expanded into several directions, including modeling and forecasting. For a recent review on some of these streams of research see \cite{petro:20}. 
A well-accepted approach for modeling the adoption over time of RETs relies on innovation diffusion models or growth curves. Starting from the contributions given by Marchetti at the IIASA in the '70s and '80s (\cite{marchetti:79}, \cite{marchetti:80}) this methodology hypothezises that energy sources are similar to commercial products that may be accepted by consumers or not. Consumers may show a different attitude in this sense, choosing whether being `pioneers' or waiting to see the decisions of others. As hypothesized by \cite{marchetti:80}, energy consumption dynamics give rise to collective learning processes based on information sharing and imitation of others' behavior, which are best described through growth curves, like the logistic equation. From these first essential ideas, the contributions have developed in more recent times, to capture the complexity of energy system dynamics (see for instance \cite{guseo:07},  \cite{meade:15}, \cite{rao:10}), with a special focus on RETs. \cite{rao:10} observed that, despite some common traits with other technologies and products, the inherent characteristics of RETs, such as low load factor, need for energy storage, small size, high upfront costs, hinder their integration into existing power systems and this explains the support provided to renewables in the form of market incentives, like feed-in tariffs. 
To account for this aspect, some works have focused on measuring the effect of incentives on diffusion, by employing a successful extension of the well-known Bass Model (\cite{bass:69}), the Generalized Bass Model by \cite{bass:94}. In \cite{guidolin:10} a cross-country analysis of PV diffusion was performed, highlighting the ability of the GBM in capturing the effect of market incentives in national adoption patterns.  The use of the GBM also allowed a more accurate forecasting procedure through an efficient description of the nonlinear trajectory, which, in the case of renewables is often far from being smooth. In \cite{DV:11} a similar analysis was conducted on the diffusion of wind power, also performing an out-of-sample forecasting evaluation with a comparison among concurrent models. 
\cite{bunea:20} upgraded and extended previous contributions on PV diffusion, by considering the 26 countries that  contributed to 99\% of cumulative PV adoptions during 1992-2016. They confirmed the lack of support by the media system and showed that public incentives were often not efficiently designed in a long term perspective, with a kind of `addiction to incentive' phenomenon. \\
The competitive disadvantage characterizing RETs has recently stimulated some research contributions trying to study adoption patterns within a competitive environment, by accounting for the influence of incumbent sources for electricity provision, namely nuclear, gas and coal. This led to employ multivariate diffusion models in duopolistic conditions. 
In \cite{guidolin:16} the case of Germany's energy transition was studied, by modeling the competitive strength of solar and wind, jointly considered, towards nuclear power. The results of this study provided a measure for the widespread belief of the Germans towards the energy transition. \cite{furlan:18} considered the competition between traditional sources and renewables for the US, Europe, China and India. 
\cite{guidolin:19} studied the case of Australia by focusing on the relationship, either competitive or collaborative, between coal, gas and renewables. A specific finding of this analysis concerned the role of gas as a `transition mean', given its competitive role towards coal, while aiding the uptake of renewables.
The study proposed in this paper aims at providing a further insight within this branch of literature. 

\section{Diffusion models: imitation, competition, collaboration}
\label{sect:3}
The use of innovation diffusion models provides significant advantages over traditional time series models, or ``black-box'' methods widely used in data analysis, thanks to their flexibility and interpretability. 
A well-known model to analyze market penetration of a single product or technology, also in the energy context, is the Bass model, in both its standard and generalized form. 
The Bass model, BM, describes the temporal  development of a univariate diffusion process, where the hazard of becoming an adopter for a single innovation or product, is the sum of two components, one reflecting adoptions resulting from the \textit{external} or mass media pressures, the other one reflecting adoption following social or collective learning resulting from agents' spontaneous, or \textit{internal}, social interactions and related
spread of information.\\
The formal representation of the BM is a first order differential equation
\begin{equation}\label{011}
z'(t)=\left(p+q\frac{z(t)}{m}\right)(m-z(t)),
\end{equation}
where the variation over time of consumption, $z'(t)$, is proportional to the residual market, $m-z(t)$, with $m$ the constant market potential and $z(t)$ cumulative adoptions at time $t$. Parameter $p$ represents the effect of the external information,  while $q$ is the internal coefficient, whose influence is modulated over time by the ratio $z(t)/m$.\\
Multivariate generalizations of the BM, aiming to describe  diffusion processes where more than one innovation concur, have been proposed by several authors, for instance \cite{guseo:14}, \cite{guseo:15}, \cite{krishnan:00}, \cite{laciana:14} and \cite{savin:05}. 

In particular, this paper employs the general model for a diachronic duopolistic competition proposed in \cite{guseo:14}, called \textit{unbalanced competition and regime change diachronic model}, UCRCD. 
The UCRCD model postulates a diffusion process characterized by two phases: an initial phase where only one (energy production) technology is available, and therefore energy production essentially represents a monopolistic regime, and a later stage where a competition regime starts due to the entrance of a concurring technology. Borrowing a typical game-theory language, the first technology, may be termed the \textit{incumbent}, while the second, entering the market at a second stage, may be referred as the \textit{entrant}. Given these different phases, the market potential may have different levels: $m_a$, the market potential of the incumbent in the monopolistic phase, and $m_c$, the market potential under competition. The residual market $m-z(t)$ is assumed to be shared, where $z(t)=z_1(t)+z_2(t)$ are overall cumulative adoptions.
The second technology enters the market at time $t=c_2$ with $c_2>0$.\\
The model is described by a system of two differential equations where $z'_1(t)$ and $z'_2(t)$ indicate instantaneous adoptions of the first and of the second technology, respectively, and $I_A$ is the indicator function of time interval $A$,
\begin{eqnarray} \label{2.1}
z'_1(t) &=& m\Bigg\{\left[p_{1a}+q_{1a}\frac{z(t)}{m}\right](1-I_{t>{c_2}})\nonumber\\
&&+\left[p_{1c}+(q_{1c}+\delta)\frac{z_1(t)}{m}+q_{1c}\frac{z_2(t)}{m}\right]I_{t>{c_2}}\Bigg\}\left[1-\frac{z(t)}{m}\right] ,\\
z'_2(t) &=& m\left[p_{2}+(q_{2}-\gamma)\frac{z_1(t)}{m}+q_{2}\frac{z_2(t)}{m}\right]\left[1-\frac{z(t)}{m}\right]I_{t>{c_2}},\nonumber\\
m &=& m_a(1-I_{t>{c_2}})+m_cI_{t>{c_2}}\nonumber\\
z(t) &=& z_1(t)+z_2(t)I_{t>{c_2}}.\nonumber
\end{eqnarray}
In the monopolistic phase, $t \leq c_2$, the trajectory of the incumbent, $z'_1(t)$, is described according to a standard Bass model with parameters $p_{1a}$, $q_{1a}$, and $m_a$.
In the competition phase, $t>c_2$, competitors influence each other according to the full UCRCD model.
The incumbent is characterized by new parameters: the innovation coefficient under competition, $p_{1c}$, the \textit{within} imitation coefficient $q_{1c}+\delta$, reflecting learning of potential adopters from adopters of technology 1, and the \textit{cross} imitation one, $q_{1c}$, which is powered by $z_2/m$ and measures the effect of the diffusion of the entrant on the incumbent.
The entrant has three corresponding parameters: the innovation coefficient $p_2$, the \textit{within} imitation coefficient $q_2$, modulating internal growth through the ratio $z_2/m$ and the \textit{cross} imitation coefficient $q_2-\gamma$, which measures the effect, of the incumbent. Typically parameters $\delta$ and $\gamma$ are assumed to be different,  and the model is called \textit{unrestricted} UCRCD.
If the restriction $\delta=\gamma$ applies, the model takes a reduced form, called \textit{standard} UCRCD ( \cite{guseo:14}), and a symmetric behavior between the two competitors is assumed. Unlike the unrestricted model, the restricted one has a full closed-form solution.\\

The relationship between the incumbent, 1, and the entrant, 2, is measured through the cross imitation coefficients $q_{1c}$ and $(q_2-\gamma)$, which may be either positive or negative, producing both competitive and collaborative dynamics. Table 1 summarizes these possible scenarios (\cite{guidolin:19}).\\

\begin{table}
\label{tab1}
\centering
{\normalsize 
\begin{center}
\begin{tabular}{c| c| c}
  ${q_{1c}}$ &${q_2-\gamma}$& interpretation\\
  \hline\hline
  negative & negative & full competition  \\
  negative & positive & 2 competes with 1, 1 collaborates with 2 \\
  positive & negative  & 2 collaborates with 1, 1 competes with 2 \\
  positive & positive & full collaboration \\
   
\hline\hline
   \end{tabular}
\end{center}
} 
\caption{The UCRCD model: cross imitation parameters and their `competition-collaboration' interpretation}
\end{table}

\subsection{Statistical inference and estimation}
\label{sect:3.1}
The statistical implementation of the UCRCD models is based on nonlinear least squares, NLS, (\cite{seber:89}). The structure of a nonlinear regression model may be considered
\begin{equation}
\label{4}
w(t)=\eta(\beta,t)+\varepsilon(t),
\end{equation}
where $w(t)$ is the observed response, $\eta(\beta,t)$ is the deterministic component describing instantaneous or cumulative processes, depending on parameter vector $\beta$  and time t, and $\varepsilon(t)$ is a residual term, not necessarily independent identically distributed (i.i.d.). \\
Model goodness-of-fit may be evaluated through the $R^2$ index, while model selection between an extended model, $m_2$, and a nested one, $m_1$, may be evaluated through a
squared multiple partial correlation coefficient $\tilde{R}^2$ (lying in the interval $[0;1]$), 
\begin{equation}\label{5}
\tilde{R}^2 = (R_{m_2}^2 - R_{m_1}^2)/(1 - R_{m_1}^2),
\end{equation}
where $R_{m_i}^2, \;i=1,2$ is the standard determination index of model $m_i$.\\
The $\tilde{R}^2$ coefficient has a monotone relationship with the $F$-ratio, i.e.,
\begin{equation}\label{6}
F = [\tilde{R}^2 (n-v)]/[(1 - \tilde{R}^2) u],
\end{equation}
where $n$ is the number of observations, $v$ the number of parameters of the extended model $m_2$, and $u$ the incremental number of
parameters from $m_1$ to $m_2$.\\

%
%
%
%
%
%

\section{Analysis of selected case-studies}
\label{sect:4}
\subsection{Dataset description}
The data considered in this study refer to the time series of consumption of natural gas and renewables (solar PV and wind jointly studied) for the period 1965-2019, according to the \cite{BP:20}. 
Figure \ref{fig1} displays the observed data for each of the countries analyzed. Indeed, the scale of the processes depends on size, population and energy demand of each country, so large and obvious differences may be easily seen in the figure. However, the focus of the analysis has been the dynamics of the diffusion processes underlying gas and RETs consumption, and their interplay, which we presume are tractable with the same modeling approach and comparable. \\As a general pattern, characterizing all countries, RETs present a strongly growing trend, after an initial phase where the data show a flat behavior, due to the difficulties connected with the adoption of renewables, then overcome with suitable incentive measures. This exponential growth has undergone a slowdown in most recent years in just two cases, Italy and Spain, where the `addiction to incentive' phenomenon suggested by \cite{bunea:20} may have been especially strong. Concerning natural gas data, a stronger variability appears among countries, although a substantially positive trend may be observed, with the remarkable exceptions of China on the one hand, with a strongly exponential growth, and Denmark on the other, where the trajectory is clearly in its declining phase. 

\subsection{Results}
\begin{figure}
\includegraphics[width= 6in]{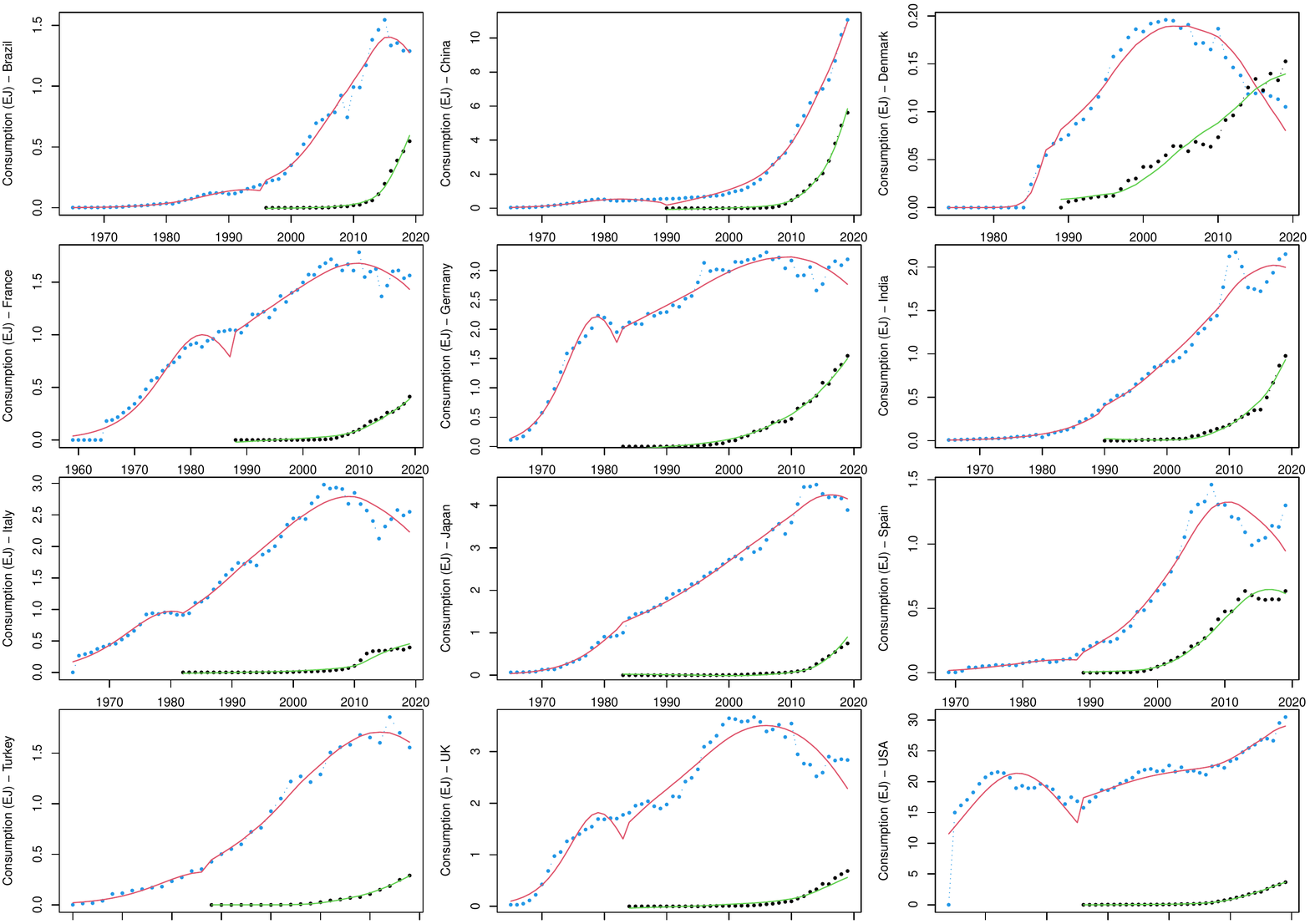}
\caption{The dynamic relationship between consumption (measured in ExaJoules) of natural gas (blue dots: observed, red line: predicted) and renewables (black dots: observed, green line: predicted) in the 12 countries considered, 1965-2019, as resulting from the fit of the  UCRCD model.} 
\label{fig2}
\end{figure}

This section describes the results of the application of the UCRCD model to the 12 selected countries. 
Parameter estimates of the model for $t> c_2$ are outlined in Table 2 and model fitting is illustrated in Figure \ref{fig2}. 
As a general finding, the model obtains an extremely satisfactory result in terms
of goodness-of-fit and significance of parameters. This confirms the existence of a significant dynamic relationship between gas and renewables in all the cases considered, which was not sure \textit{a priori}. While for most countries considered the model selection procedure led to prefer the restricted UCRCD model, in view of its parsimony, for three countries namely Denmark, Italy and the USA, the unrestricted version, with $\delta \neq \gamma$, proved better (in italics in the table).
The only coefficient always resulting non-significant was $p_2$, the external coefficient for renewables, and therefore not displayed within results. This fact is consistent with the flat behavior of data observed before and with previous findings on the weak role played by the external component in the adoption of RETs, which justified the need of \textit{ad-hoc} incentive measures to stimulate market growth (\cite{guidolin:10}, \cite{bunea:20}). 
Parameters $p_{1c}$ and $(q_{1c}+\delta)$, referred to the trajectory of gas, are significant and positive, confirming an essentially growing trend, although the magnitude of parameter $q_2$ in all the countries, calls the attention on the more intense growth of RETs. \\
The most interesting insights clearly come from sign and magnitude of cross imitation parameters $q_{1c}$ and $(q_2-\gamma)$. 
For all countries, with the noteworthy exception of the USA, the cross imitation coefficient $q_{1c}$ is always negative and significant, showing a competitive pressure of RETs over gas. So RETs appear as mature technologies able to compete in the energy market. 
This is not the case only in the USA, where this coefficient is positive and disproportionately high, $q_{1c}= 1.34$, suggesting that gas plays a dominant position in this market, and the growth of renewables is apparently reinforcing this situation. This is also confirmed by the fact that parameter $(q_2-\gamma)= 0$ in USA, so that gas appears to have no role in the growth of RETs, not competing nor collaborating with them. \\
Concerning the effect of gas towards RETs, in eight countries $(q_2-\gamma)$ is positive, and therefore a collaboration dynamics is detected. However, the magnitude of this effect is much smaller than the competitive effect of RETs 
For an overview of the nature of the relationship between gas and RETs,  Table 3 summarizes the results as emerged from the fit of the UCRCD model. 
Notably, in four countries, namely Denmark, India, Japan and Turkey, the interplay between RETs and gas is of \textit{pure competition}, since $(q_2-\gamma)$ is negative. Interestingly, in these cases, the data of gas have started to decrease. This is especially evident in Denmark, which may be taken as a paradigmatic example of competition and substitution between technologies. 

As an overall remark arising from these results, we may infer that an interplay of pure competition emerges, whenever RETs are growing and gas is declining. 
In other cases, where gas is still growing, although slightly, the model captures a competition-collaboration pattern, where gas is found to have the role of a \textit{transition mean} or \textit{bridging technology}. 
The only exception to this situation is represented by the USA, whose peculiar results, contradicting all other findings, require a specific investigation.

\begin{table}
\normalsize
  \centering 
 \label{tab2}
 \begin{tabular}{@{}l@{\hskip 1.5cm}S@{\hskip 0.5cm}S@{\hskip 0.5cm}S@{\hskip 0.5cm}S@{\hskip 0.5cm}S@{\hskip 0.5cm}S@{\hskip 0.5cm}S@{\hskip 0.5cm}S}  
\hline\hline
 Country & $m_c$ &$p_{1c}$ &${(q_{1c}+\delta)}$&$q_{1c}$	&$q_{2}$ &{$(q_{2}-\gamma)$}&$\delta$ &$\gamma$\\
\hline

Brazil &61 & 0.003&0.12 &-0.29&0.41&0.002&0.41& \\

China &2429 &0.000&0.13&-0.05&0.2&0.010&0.19&\\

\textit{Denmark} &	10&	0.007&	0.11&-0.19&0.22&-0.010 &0.30 &0.23\\

France &109& 0.009&0.04&-0.18&0.23&0.002&0.23&\\

Germany& 409&0.004&0.03 &-0.10&0.14 &0.003&0.13&\\

India  & 158& 0.002&0.08&-0.16&	0.24&-0.001 &0.24&\\

\textit{Italy} & 131& 0.006 & 0.07& -0.16&0.34& 0.001&0.23& 0.34\\

Japan & 532& 0.002&0.04&-0.27&0.32&-0.001&0.32&\\

Spain &48&	0.002&0.14&-0.09& 0.24&0.004&0.23&\\

Turkey& 52&0.007&0.13&-0.32&0.45&-0.0002&	0.45\\

UK  &153&	0.009&0.07& -0.33&	0.40&0.001&0.40&\\

\textit{USA}	& 1257&0.013&0.03&1.34 &0.39 &-0.000&-1.3& 0.39\\
 
\hline\hline

\end{tabular}
 \caption{The dynamic relationship between consumption of natural gas and renewables in the 12 countries considered, 1965-2019. Parameter estimates of the UCRCD model for the competition phase ($t > c_2$).}
\end{table}

\begin{table}
\normalsize
  \centering 
 \label{tab3}
 \begin{tabular}{l| c| l}
\hline\hline
 Country & RETs vs gas & gas vs RETs\\
\hline

Brazil & competition& collaboration\\

China &competition& collaboration\\

Denmark &competition& competition \\

France &competition& collaboration\\

Germany& competition& collaboration\\

India  & competition& competition\\

Italy & competition& collaboration\\

Japan & competition& competition\\

Spain &competition& collaboration \\

Turkey& competition& competition\\

UK  &competition& collaboration \\

USA	& collaboration& no effect \\
 
\hline\hline

\end{tabular}
\caption{Interplay between gas and RETs}
\end{table}

\section{Discussion}
\label{sect:5}
Our analyses from the pre-pandemic epoch show a
clear competitive effect exerted by renewables, that seem to
follow a robust and somehow independent growth path. On the other hand, natural gas is found to have in most cases a `bridging technology' role. 
However, the lack of persistent support to renewables from the public and the media system, as documented by
the negligible values of the external coefficients, as well as the unavoidable
uncertainty due to the pandemic, indicate that the market for these technologies will still continue to
require important investments to achieve a sustained path allowing to meet the ``climate targets''.

\section*{Acknowledgments}
{This research has been partially funded by the grant \textbf{BIRD188753/18} of the University of Padua, Italy}

\section*{Appendix} 
We report here full parameter estimates of the UCRCD model for the 12 countries considered. 
\begin{table}[h]
\footnotesize
  \centering 
 \label{tab3}
 \begin{tabular}{lcccccc|c}  
\hline\hline
country & parameter & estimate & s.e. & \hbox{lower} c.i. & \hbox{upper} c.i. & $p$-value & ${R^2}$\\
\hline
&${m_c}$& 60.9 &8.24&44.7&77.7&$<0.0001$&0.999824\\
 \textit{Brazil}&$p_{1c}$& 0.003&0.0004&0.002&0.004&$<0.0001$\\
&$p_{2c}$& -0.00080&0.00003&-0.00083&-0.00067&0.83\\
&$q_{1c}$& -0.291&0.0376&-0.365&-0.217&$<0.0001$\\
&$q_{2c}$& 0.414&0.0367&0.342&0.486&$<0.0001$\\
&${\delta}$& 0.412&0.0391&0.336&0.489&$<0.0001$\\
\hline
&$m_c$&2429.4&1941.7&-1376.3&6235.1&0.216&0.999398\\
\textit{China}&$p_{1c}$&0.00004&0.00003&-0.00002&0.00011&0.186\\
&$p_{2c}$&-0.00003&0.00004&-0.00011&0.00005&0.445\\
&$q_{1c}$&-0.0595&0.0129&-0.0848&-0.0342&$<0.0001$\\
&$q_{2c}$&0.2052&0.01202&0.1816&0.2287&$<0.0001$\\
&$\delta$&0.1934&0.0146&0.1648&0.2221&$<0.0001$\\
\hline
&$m_c$&10.1&1.3&7.5&12.6&$<0.0001$&0.9996415\\
\textit{Denmark}&$p_{1c}$&0.0074&0.001&0.0054&0.0093&$<0.0001$\\
&$p_{2c}$&0.0009&0.0007&-0.0004&0.0023&0.176\\
&$q_{1c}$&-0.1910&0.032&-0.2538&-0.1282&$<0.0001$\\
&$q_{2c}$&0.2225&0.065&0.0948&0.3502&0.0012\\
&$\delta$&0.3035&0.0345&0.2358&0.3711&$<0.0001$\\
&$\gamma$&0.2326&0.0771&0.0816&0.3837&0.0038\\
\hline
&$m_c$&109.1&9.8&90.2&128.3&$<0.0001$&0.9998478\\
\textit{France}&$p_{1c}$&0.0091&0.0007&0.0078&0.0105&$<0.0001$\\
&$p_2$c&-0.0002&0.0002&-0.0007&0.0002&0.365\\
&$q_{1c}$&-0.1870&0.0276&-2.411&-0.1329&$<0.0001$\\
&$q_{2c}$&0.2346&0.0264&0.1828&0.2865&$<0.0001$\\
&$\delta$&0.2320&0.0274&0.1783&0.2858&$<0.0001$\\
\hline
&$m_c$&409.5&90.1&232.9&586.1&$<0.0001$&0.9999245\\
\textit{Germany}&$p_{1c}$&0.0048&0.0010&0.0029&0.0067&$<0.0001$\\
&$p_{2c}$&-0.0002&0.0001&-0.0004&0.0001&0.255\\
&$q_{1c}$&-0.1055&0.0103&-0.1257&-0.0852&$<0.0001$\\
&$q_{2c}$&0.1403&0.0113&0.1181&0.1625&$<0.0001$\\
&$\delta$&0.1366&0.0119&0.1134&0.1599&$<0.0001$\\
\hline
&$m_c$&158.9&44.3&72.2&242.8&0.0007&0.999748\\
\textit{India}&$p_{1c}$&0.0024&0.0006&0.0013&0.0034&$<0.0001$\\
&$p_{2c}$&0.00005&0.00026&-0.00045&0.00056&0.845\\
&$q_{1c}$&-0.1621&0.0306&-0.2222&-0.1021&$<0.0001$\\
&$q_{2c}$&0.2455&0.0299&0.1870&0.3041&$<0.0001$\\
&$\delta$&0.2468&0.0330&0.1822&0.3115&$<0.0001$\\
\hline\hline
\end{tabular}
\end{table}

\begin{table}
\footnotesize
  \centering 
 \label{tab3}
 \begin{tabular}{lcccccc|c}  
\hline\hline
country & parameter & estimate & s.e. & \hbox{lower} c.i. & \hbox{upper} c.i. & $p$-value & ${R^2}$\\
\hline
&$m_c$&131.9&13.4&105.7&158.1&$<0.0001$&0.9999301\\
\textit{Italy}&$p_{1c}$&0.0067&0.0005&0.0057&0.0077&$<0.0001$\\
&$p_{2c}$&-0.0001&0.0003&-0.0007&0.0005&0.716\\
&$q_{1c}$&-0.1646&0.2203&-0.5965&0.2672&0.457\\
&$q_{2c}$&0.3426&0.0875&0.1710&0.5142&0.0002\\
&$\delta$&0.2367&0.2159&-0.1865&0.6599&0.277\\
&$\gamma$&0.3414&0.0886&0.1678&0.5150&0.0002\\
\hline
&$m_c$&532.0&128.1&280.9&783.1&$<0.0001$&0.9998761\\
\textit{Japan}&$p_{1c}$&0.0023&0.0005&0.0013&0.0032&$<0.0001$\\
&$p_{2c}$&0.00005&69.5225&-0.00009&0.00018&0.511\\
&$q_{1c}$&-0.2762&0.0310&-0.0337&-0.2154&$<0.0001$\\
&$q_{2c}$&0.3236&0.0303&0.2643&0.3829&$<0.0001$\\
&$\delta$&0.3247&0.0309&0.2641&0.3854&$<0.0001$\\
\hline
&$m_c$&48.4&1.8&44.8&52.0&$<0.0001$&0.9994917\\
\textit{Spain}&$p_{1c}$&0.0028&0.0007&0.0015&0.0041&$<0.0001$\\
&$p_{2c}$&0.0001&0.0007&-0.0012&0.0014&0.882\\
&$q_{1c}$&-0.0963&0.2760&-0.1504&-0.0422&0.0009\\
&$q_{2c}$&0.2429&0,0265&0.1910&0.2949&$<0.0001$\\
&$\delta$&0.2384&0.0327&0.1744&0.3025&$<0.0001$\\
\hline
&$m_c$&52.8&3.1&46.8&58.8&$<0.0001$&0.9999621\\
\textit{Turkey}&$p_{1c}$&0.0075&0.0005&0.0066&0.0084&$<0.0001$\\
&$p_{2c}$&0.00002&0.0004&-0.0009&0.0009&0.961\\
&$q_{1c}$&-0.3234&0.0819&-0.4838&-0.1629&0.0003\\
&$q_{2c}$&0.4521&0.0786&0.2981&0.6061&$<0.0001$\\
&$\delta$&0.4523&0.0815&0.2926&0.6120&$<0.0001$\\
\hline
&$m_c$&153.9&6.15&141.8&165.9&$<0.0001$&0.9997931\\
\textit{UK}&$p_{1c}$&0.0099&0.0005&0.0089&0.0109&$<0.0001$\\
&$p_{2c}$&-0.0003&0.0005&-0.0013&0.0008&0.632\\
&$q_{1c}$&-0.3306&0.0906&-0.5081&-0.1531&0.0005\\
&$q_{2c}$&0.4035&0.0885&0.2302&0.5769&$<0.0001$\\
&$\delta$&0.4016&0.0899&0.2255&0.5778&$<0.0001$\\
\hline
&$m_c$&1257.2&118.0&1026.0&1488.5&$<0.0001$&0.9998681\\
\textit{USA}&$p_{1c}$&0.0136&0.0012&0.0112&0.0160&$<0.0001$\\
&$p_{2c}$&0.000001&0.0002&-0.0004&0.0004&0.995\\
&$q_{1c}$&1.3458&0.3790&0.6030&2.0887&0.0007\\
&$q_{2c}$&0.3984&0.0755&0.2504&0.5465&$<0.0001$\\
&$\delta$&-1.3107&0.3766&-2.0489&-0.5725&0.0009\\
&$\gamma$&0.3987&0.0762&0.2493&0.5481&$<0.0001$\\
\hline\hline
\end{tabular}
\end{table}

\end{document}